\documentstyle[11pt]{article}

\title{\bf The intercept of symmetric multigluon configurations in the
variational approach}
\author{N. Armesto and M. A. Braun$^*$\\
{\it  Departamento de F\'{\i}sica de Part\'{\i}culas,}\\
{\it Universidade de Santiago de Compostela,}\\
{\it 15706--Santiago de Compostela, Spain}}
\date{}
\def\beq{\begin{equation}}
\def\eeq{\end{equation}}
\def\noi{\noindent}
\begin{document}
\maketitle
\medskip
\vspace{0.5cm}
\centerline{{\large {\bf Abstract}}}
\begin{quotation}
To calculate the intercept of the multigluon system in a
symmetric spatial configuration a variational method is developed based on a
complete system of one-gluon functions. The
method is applied to two- and three- gluon cases to compare with the known
results. The convergence turns out rather slow. Ways to improve results are
discussed.
\end{quotation}
\vspace{8.5 cm}

\noi{\large October 1994}

\noi{\Large\bf US-FT/9-94}

\vspace{0.5cm}
\noi$^*$ On leave of absence from the Department of High Energy
Physics, University of St. Petersburg, 198904 St. Petersburg, Russia.

 \newpage

\section{Introduction}

\ \ \ \ Much attention has recently been devoted to the perturbative "hard", or
BFKL, pomeron \mbox{[1]}, especially in relation to the study of the small $x$
behaviour
 of the deep inelastic scattering structure functions (see a recent review
in [2]). In application to soft phenomena, the value of the pomeron
intercept is of principal importance. For the BFKL pomeron it is
considerably above unity:
\[ \alpha_{BFKL}(0)=1-(3\alpha_{s}/\pi)E_{0},\]
where the "energy" $E_{0}$ is equal $-4\ln 2$, and $\alpha_{s}$ is the (fixed)
QCD coupling constant [1]. However the BFKL pomeron is only the simplest of
the family of pomerons (with
 positive signature) and odderons (with negative signature) formed in a
system of $n$ interacting reggeized gluons. Should some of them also result
supercritical (with the intercept above unity), the study of the high energy
behaviour of the QCD cross-sections would require summation of an arbitrary
number of exchanges of all these supercritical pomerons and odderons.

It is extraordinarily difficult to obtain an explicit solution or an exact
energy value for $n>2$ interacting gluons (see, however, some ideas in [3]).
For $n=3$ (odderon) application of conformal symmetry allows to reduce
calculations to an one dimensional problem [4]. Variational treatment
with a simple trial wave function
 then gives an intercept also above unity, although lower than for the
pomeron [5]:
\[\alpha_{odd}(0)=1-(3\alpha_{s}/\pi)E_{odd},\ E_{odd}<-0.37.\]
For larger number $n$ of pomerons different crude approximations lead to wildly
different estimates for the intercept. Developing the interaction in powers
of the conformal group Casimir operator  an intercept linearly rising with
$n$ is obtained [6] . On the other hand,
  the Hartree-Fock approximation, presumably valid for large $n$, gives
intercepts less than one and linearly falling with $n$ [7]. In this
approximation the energy is
$\epsilon$ rising  linearly with $n$:
\[ E_{n}=(n/2)*0.959...\]

Evidently negative energies, as witnessed by the pomeron and odderon, result
from correlations. In absence of explicit solutions for $n>2$ gluons and in
view of great technical difficulties in working with conformally
invariant wave functions, evident already for the odderon case, the only
realistic approach for $n>3$ gluons seems to be a direct variational one,
based on some complete and simple basis of functions. Taking a finite number
$N$ of these, one then computes a finite energy matrix. The intercept (minus
one) is   found as its smallest eigenvalue (ground state) with an opposite
sign. With $N$ growing,  the ground state energy goes down, so that one always
obtains its upper limit. As $N\rightarrow\infty$ one is sure to obtain an
exact value, provided the basic functions form a complete set. One can have
some idea of the precision for a given $N$ by calculating the energy for the
two-gluon case where its exact value is known.

In this paper we give an outline of such a variational method aimed at
calculating of the intercept of the system of an arbitrary number of reggeized
gluons in a symmetric configuration both in the ordinary space and in the
colour space.
The basic functions were chosen as harmonic oscillator
functions in the variable $z=\ln r^{2}$ (radials) multiplied by azymuthal
functions for given angular momenta $l=0,\pm1,\pm2,...$. They retain some
of the conformal invariance corresponding to the substitution $r\rightarrow
1/r$. To see the convergence, we studied the two-gluon (the BFKL pomeron)
and three-gluon (odderon) cases, where the exact energy value for the former
and its upper limit for the latter are known.  It turns out that the
convergence is rather slow. In the present series of calculations up to 201
basic states have been included for the pomeron  and up to 1335 basic states
for the odderon.
 With these
basic states the resulting upper limit achieved for the two-gluon ground state
energy (the BFKL pomeron) is $-1.032$, which means
$\sim 53\%$ of the exact correlation energy.
 The upper limit obtained for the odderon ground state energy
is
$+0.331$. This limit lies considerably above the one obtained in [5] with a
conformally invariant ansatz. From the known BFKL intercept we can study the
dependence of the calculated energy on the number of basic states taken into
account. Applying a similar fit to the odderon case and extrapolating to
infinite number of states included gives an estimate
\[ E_{odd}<-0.3\div -0.6,\]
in agreement with [5].

\section{Variational calculation of the ground state energy for $n$ reggeized
gluons}

\ \ \ \ As shown in [8] the transverse space and colour wave function $\psi$
of
$n$ reggeized gluons in a colourless state satisfies a Schr\"odinger-like
equation
\beq
H\psi=E\prod_{i=1}^{n}p_{i}^{2}\psi,
\eeq
where $p_{i}$ is the momentum of the $i$-th gluon. The Hamiltonian $H$ is
given by a sum of pair terms
\beq
H=-(1/6)\sum_{i<k}T_{i}T_{k}H_{ik}.
\eeq
Here $T_{i}$ is the colour vector of the $i$-th gluon. In a colourless state
\beq
\sum_{i=1}^{n}T_{i}=0.
\eeq
The pair Hamiltonian $H_{ik}$ acts on the wave function according to
\beq
H_{ik}\psi=\prod_{j=1}^{n}p_{j}^{2}(\ln p_{i}^{2}p_{k}^{2}+4{\bf C})\psi
+\prod_{j=1,j\neq i,k}^{n}p_{j}^{2}(p_{i}^{2}\ln (r_{ik}^{2}/4)\,p_{k}^{2}
+(i\leftrightarrow k))\psi+2(p_{i}+p_{k})^{2}\psi(r_{ik}=0),
\eeq
where $r_{ik}=r_{i}-r_{k}$ is the (transversal) distance between the gluons;
$\bf C$ is the Euler constant.
The energy eigenvalue $E$ is related to the intercept
by
\beq
\alpha(0)=1-(3\alpha_{s}/\pi)E,
\eeq
so that the rightmost singularity in the complex angular momentum plane
corresponds to the ground state. As mentioned, for $n=2$ the solution of (1) is
the BFKL pomeron with
\beq
E_{0}=-4\ln 2,\ \alpha_{BFKL}(0)=1+(\alpha_{s}/\pi)12\ln 2.
\eeq
No explicit solution has been found for $n>2$.

The solution of (1) may evidently be found by a variational approach,
searching
 the minimum value of the functional
\beq
\Phi=\int\prod d^{2}p_{i}\psi^{\ast}H\psi\equiv\sum_{i<k}\Phi_{ik},
\eeq
with the normalization condition
\beq
\int\prod d^{2}p_{i}\psi^{\ast}\prod_{i=1}^{n}p_{i}^{2}\psi=1.
\eeq

The BFKL pomeron and odderon states are symmetric both in colour and space
variables. For more gluons solutions do not presumably possess such simple
symmetry properties, since  Eq. (1) mixes colour and space variables. Among
various solutions there certainly are those which correspond to a number of
BFKL pomerons weakly interacting with each other (multipomeron cuts in the
old Regge-pomeron theory). These solutions are basically symmetric neither in
space nor in colour variables. We are interested in solutions of a different
type, which do not reduce to BFKL pomerons and represent new pomerons with a
higher intrinsic colour.
While exact symmetry properties of these new states are not evident, one
expects that in the variational approach they can well be approximated by
states symmetric both in colour and space variables, similar to the simplest
BFKL pomeron and odderon states. In any case, the energy value obtained with
such a choice will give an upper limit for the exact one.  According to the
colour cluster separation property discussed in [7], these new pomerons (and
odderons) will also appear as subsystems for still larger number of gluons,
that is, will take part in multipomeron exchanges. Such a picture was
phenomenologically introduced in a model of colour string fusion [9], where
it was supposed that supercritical pomerons exist for arbitrary high intrinsic
colour.

As noticed in [4], for a wave function
symmetric both in colour and space variables, instead of the full functional
$\Phi$ one can use any of its pair terms $\Phi_{ik}$ in (7) with $i$ and $k$
fixed. Moreover, for a symmetric colour wave function,
\beq
<T_{i}T_{k}>=(2/n(n-1))\sum_{i<k}T_{i}T_{k}=-3/(n-1),
\eeq
where (3) and $T^{2}=3$ have been used. As  a result, the energy $E$ may be
found via the minimal value of the functional in only spatial variables
\beq
{\cal E}=(1/2)\int\prod_{i=1}^{n}d^{2}p_{i}\psi^{\ast}H_{12}\psi,
\eeq
where the Hamiltonian $H_{12}$ is defined by (4) with $ik=12$ and the
function $\psi$ should satisfy (8). The energy of the whole system of $n$
gluons is determined by the minimal value  $\epsilon$ of $\cal E$  according
to \beq
E_{n}=(n/2)\epsilon_{n}.
\eeq
Actually the operator in $\cal E$ does not depend on $n$. The dependence on
$n$ enters only from extra arguments in $\psi$ through the requirement of the
symmetry in all arguments. The space of trial wave
functions  then gets smaller with rising $n$, wherefrom one obtains
\beq
\epsilon_{n+1}\geq\epsilon_{n}
\eeq and in particular $\epsilon_{n}\geq\epsilon_{2}$. This gives a crude
lower limit on the energy and an upper one on the intercept for $n$ gluons in a
symmetric state
\beq
E_{n}\geq (n/2)E_{0}.
\eeq
This limit was obtained in [4] for the odderon $n=3$.

For a large number of  reggeized gluons one can hopefully apply the
Hartree-Fock approximation and seek for the minimum of the functional $\cal
E$ on functions $\psi$ which factorize into a product of individual gluon wave
functions. For the ground state, with $\psi$ symmetric in all gluons, all
individual gluon wave functions should evidently be the same. As calculated
in [7], the obtained minimal Hartree-Fock value is $\epsilon=0.959...$,
so that the Hartree-Fock energy  of the symmetric $n$ gluon
state is positive and grows linearly with $n$, as stated in the Introduction.
It means that there is little hope to expect supercritical pomerons composed
of a large number of gluons. It also means that supercritical pomerons are
formed exclusively due to correlations in the wave function.

For a finite number $n$ of gluons one can expand the symmetric spatial wave
function in a sum of products of individual gluon functions:
\beq
\psi(r_{1},...,
r_{n})=\sum_{\alpha_{1},...,\alpha_{n}}c_{\alpha_{1},...,\alpha_{n}}\prod
_{i=1}^{n}
\psi_{\alpha_{i}}(r_{i}),
\eeq
where the one-gluon functions $\psi_{\alpha}(r_{i})$ form a discrete
complete set and are orthonormalized according to (8):
\beq
\int d^{2}r\psi_{\alpha}^{\ast}p^{2}\psi_{\alpha'}=\delta_{\alpha,\alpha'}.
\eeq
The coefficients $c_{\alpha_{1},...,\alpha_{n}}$ have to be symmetric in all
$\alpha$'s by the requirement of the Bose symmetry and normalized according to
\beq
\sum_{\alpha_{1},...,\alpha_{n}}|c_{\alpha_{1},...,\alpha_{n}}|^{2}=1
\eeq
The two-gluon
Hamiltonian
$H_{12}$ acts nontrivially only on the wave functions for the gluons number
one and two. So the energy functional becomes
\beq
{\cal E}=\sum_{\alpha_{1},\alpha_{2},\alpha'_{1},\alpha'_{2},\alpha_{3},...,
\alpha_{n}}c^{\ast}_{\alpha_{1},\alpha_{2},\alpha_{3},...,\alpha_{n}}
c_{\alpha'_{1},\alpha'_{2},\alpha_{3},...,\alpha_{n}}{\cal
E}_{\alpha_{1},\alpha_{2},\alpha'_{1},\alpha'_{2}},
\eeq
where the matrix ${\cal E}_{\alpha_{1},\alpha_{2},\alpha'_{1},\alpha'_{2}}$
is the two-gluon energy  in the basis formed by functions $\psi_{\alpha}$.
With this matrix known, the problem of  minimization of the functional
$\cal E$ reduces to finding the minimal value of a cuadratic form, that is,
 the minimal eigenvalue of the matrix ${\cal
E}_{\alpha_{1},\alpha_{2},\alpha'_{1},\alpha'_{2}}$ considered as a matrix in
independent initial and final $n$-gluon states. The latter means that this
matrix should be multiplied by unity matrices for  the rest of the gluons
and then symmetrized in all initial and final gluons. The procedure is quite
straightforward, once the basic functions $\psi_{\alpha}$ are chosen. It
however involves a numerical evaluation of the energy matrix elements and a
diagonalization of the matrix, whose dimension is rapidly growing with the
number of gluons and the basic functions taken into account.

\section{Two-gluon energy matrix for given angular momenta}

\ \ \ \ The first task in the calculation of the energy matrix is to separate
radial and angular dependence. The basic functions depend on the azymuthal
angle
$\phi$ trivially:
\beq
\psi_{\alpha}({\bf r})=\psi_{k,l}(r)\exp il\phi,
\eeq
where $l=0,\pm1,\pm2,...$ and $k=0,1,2,...$ enumerates the radial functions.
Thus $\alpha=\{k,l\}$ is in fact a pair of indices. In the following, instead
of $r$, we shall use the variable $z=\ln r^{2}$ in most cases. In terms of
$z$ and $\phi$
\beq
p^{2}=-(4/r^{2})(\partial^{2}_{z}+(1/4)\partial^{2}_{\phi}),
\eeq
so that acting on the function $\psi_{k,l}({\bf r})$
\beq
p^{2}\psi_{k,l}({\bf r})
=-(4/r^{2})(\partial^{2}_{z}-(1/4)l^{2})\psi_{k,l}({\bf r}).
\eeq
Wave functions with different values of the angular momentum are
automatically orthogonal. For coinciding $l$  the normalization
condition for the radial functions takes the form
\beq
\int dz \psi^{\ast}_{k,l}(z)(-\partial^{2}+(1/4)l^{2})\psi_{k',l}(z)=(1/4\pi )
\delta_{kk'}.
\eeq
It reduces to the standard form for functions
\beq
\xi_{k,l}(z)=(\partial+|l|)\psi_{k,l}(z),
\eeq
which evidently satisfy
\beq
\int dz \xi^{\ast}_{k,l}(z)\xi_{k',l}(z)=(1/4\pi )
\delta_{kk'}.
\eeq
In the following we assume that the radial functions are chosen to be real.

With the angular dependence of the wave function explicitly given by (18), one
can do the azymuthal integrals in the potential energy in a straightforward
manner. Let $\alpha_{i}=\{k_{i},l_{i}\}$ and take the transition between two
gluon states
$\alpha_{1},\alpha_{2}\rightarrow\alpha_{3},\alpha_{4}$.
 Evidently the total
angular momentum is conserved so that the energy matrix elements are zero
unless
$ l_{1}+l_{2}=l_{3}+l_{4}$.
 According to (4) the potential energy consists of
two parts, the first
 part $U$ given by an essentially Coulomb interaction between the gluons and
the second one $Q$ given by a contact interaction, proportional to their
total momentum squared. Let us begin with the Coulomb part $U$. Its two terms
evidently give the same contribution due to the symmetry under the
interchange of gluons 1 and 2. So we can take only one of them and drop
the factor $1/2$. Denote
\beq
\eta_{k,l}(z)=(\partial^{2}-(1/4)l^{2})\psi_{k,l}(z).
\eeq
Then after doing the azymuthal integration
 we
obtain the following matrix elements for the potential energy $U$
\beq
U_{\alpha_{1},\alpha_{2};\alpha_{3},\alpha_{4}}=16\pi^{2}
\int dz_{1}dz_{2}\eta_{\alpha_{1}}(z_{1})\psi_{\alpha_{3}}(z_{1})
\psi_{\alpha_{2}}(z_{2})\eta_{\alpha_{4}}(z_{2})U_{l}(z_{1},z_{2}),
\eeq
where
$l=|l_{1}-l_{3}|=|l_{2}-l_{4}|$
is the angular momentum transfer and the function $U_{l}(z_{1},z_{2})$
is given by
\beq
U_{l}=-(1/l)\exp (-(l/2)(|z_{1}-z_{2}|), \ \ l\neq 0,
\eeq
and
\beq
U_{0}=\max \{z_{1},z_{2}\}.
\eeq

The contact part $Q$ involves gluonic wave functions taken at the same point,
that is, with the same $r$ and $\phi$. After performing the azymuthal
integration and integrating once by parts in the variable $z$ we obtain
\beq
Q_{\alpha_{1},\alpha_{2};\alpha_{3},\alpha_{4}}=8\pi^{2}
\int dz ((\partial +(1/2)|l_{1}+l_{2}|)\psi_{\alpha_{1}}\psi_{\alpha_{2}})
((\partial +(1/2)|l_{3}+l_{4}|)\psi_{\alpha_{3}}\psi_{\alpha_{4}}).
\eeq
One can somewhat simplify this expression by noting that
\beq
(\partial +(1/2)|l_{1}+l_{2}|)\psi_{\alpha_{1}}\psi_{\alpha_{2}}=
\xi_{\alpha_{1}}\psi_{\alpha_{2}}+\psi_{\alpha_{1}}\xi_{\alpha_{2}}
+\Delta_{12}\psi_{\alpha_{1}}\psi_{\alpha_{2}},
\eeq
where $2\Delta_{12}=|l_{1}+l_{2}|-|l_{1}|-|l_{2}|$ and similarly for the second
factor in (28). Then finally
\beq
Q_{\alpha_{1},\alpha_{2};\alpha_{3},\alpha_{4}}=8\pi^{2}
\int dz (\xi_{\alpha_{1}}\psi_{\alpha_{2}}+\psi_{\alpha_{1}}\xi_{\alpha_{2}}
+\Delta_{12}\psi_{\alpha_{1}}\psi_{\alpha_{2}})
(\xi_{\alpha_{3}}\psi_{\alpha_{4}}+\psi_{\alpha_{3}}\xi_{\alpha_{4}}
+\Delta_{34}\psi_{\alpha_{3}}\psi_{\alpha_{4}}).
\eeq

Passing to the kinetic energy given by the first terms in (4) we also note
that the two terms give the same contribution so that we can take one of them
(say, for the gluon 1) and drop the factor 1/2. For the transition of the
first gluon
$\alpha_{1}\rightarrow\alpha_{3}$
the conservation of angular momentum requires that $l_{1}=l_{3}$.
The kinetic energy is easily calculated in the momentum space. So we
transform the basic functions to the momentum space according to
\beq
\psi_{\alpha}({\bf p})=\int (d^{2}r/2\pi)\psi_{\alpha}({\bf r})\exp
  (-i{\bf pr}).
\eeq
The azymuthal integration leads to
\beq
\psi_{\alpha}({\bf p})=(-i)^{l}\exp il\phi
\,\int rdr\psi_{k,l}(z)J_{l}(pr),
\eeq
where $J_{l}$ is the Bessel function. To do the integral over $r$ it is
convenient to introduce a Fourier transform of the function $\psi$ with
respect to the variable $z$:
\beq
\psi_{k,l}(z)=\int (d\nu/\sqrt{2\pi})\phi_{k,l}(\nu)\exp i\nu z.
\eeq
Putting this representation in (33) and doing the $r$-integration we obtain
\beq
\psi_{k,l}({\bf p})=(2/p^{2})\exp il\phi
\int (d\nu/\sqrt{2\pi})f_{k,l}(\nu)p^{-2i\nu},
\eeq
with
\beq
f_{k,l}(\nu)=(-i)^{|l|}2^{2i\nu}(|l|/2+i\nu)\phi_{k,l}(\nu)
\Gamma (|l|/2+i\nu)/\Gamma(|l|/2-i\nu).
\eeq

With the gluon wave functions in the momentum space given by (34), both
radial and azymuthal integration in $\bf p$ are easily done. The final matrix
element of the kinetic energy $T$ results as
\beq
T_{\alpha_{1},\alpha_{3}}=-4\pi i\int d\nu f^{\ast}_{k_{1},l_{1}}(\nu)
(\partial/\partial\nu )f_{k_{3},l_{1}}(\nu)
\eeq
(recall that $l_{1}=l_{3}$).
The differentiation gives
\beq
(\partial/\partial\nu )f_{k_{3},l_{1}}(\nu)=f_{k_{3},l_{1}}(\nu)
(2i\ln 2+2i{\mbox Re}\, \psi (|l_{1}|/2+i\nu)+
(\partial/\partial\nu )\ln ((|l|/2+i\nu)\phi_{k,l}(\nu))).
\eeq
Correspondingly the kinetic energy matrix element separates into terms
\beq
T^{(1)}_{\alpha_{1},\alpha_{3}}=8\pi\int d\nu f^{\ast}_{k_{1},l_{1}}(\nu)
f_{k_{3},l_{1}}(\nu)(\ln 2+{\mbox Re}\, \psi (|l_{1}|/2+i\nu))
\eeq
and
\beq
 T^{(2)}_{\alpha_{1},\alpha_{3}}=-4\pi i\int d\nu
((|l_{1}|/2+i\nu)\phi_{k_{1},l_{1}}(\nu))^{\ast}(\partial/\partial\nu )
((|l_{1}|/2+i\nu)\phi_{k_{3},l_{1}}(\nu)).
\eeq

One notes that the function $((|l|/2+i\nu)\phi_{k,l}(\nu))$ is nothing but
the Fourier transform  of $\xi_{k,l}(z)$ with respect to $z$.
Correspondingly we denote it as
\beq
(|l|/2+i\nu)\phi_{k,l}(\nu)\equiv \xi_{k,l}(\nu).
\eeq
The part $T^{(2)}$ can then be written as
\beq
 T^{(2)}_{\alpha_{1},\alpha_{3}}=-4\pi i\int d\nu
\xi_{k_{1},l_{1}}(\nu)^{\ast}(\partial/\partial\nu )
\xi_{k_{3},l_{1}}(\nu)).
\eeq

The orthonormalization property (23)  transforms into the analogous
property in the $\nu$ space
\beq
\int d\nu \xi^{\ast}_{k,l}(\nu)\xi_{k',l}(\nu)=(1/4\pi)\delta_{kk'}.
\eeq
Noting that
$f^{\ast}_{k,l}(\nu)f_{k',l}(\nu)
=\xi^{\ast}_{k,l}(\nu)\xi_{k',l}(\nu)$ we observe
 that the term $\ln 2$ in (38) will add a constant $2\ln 2$ to
the energy. Separating another constant term $2\psi(1)$ we finally present
the part $T^{(1)}$ in the final form
\beq
T^{(1)}_{\alpha_{1},\alpha_{3}}=2(\ln 2+\psi(1))\delta_{\alpha_{1},\alpha_{3}}
+8\pi\int d\nu\xi^{\ast}_{\alpha_{1}}(\nu)\xi_{\alpha_{3}}(\nu)
({\mbox Re}\, \psi (|l_{1}|/2+i\nu)-\psi(1)).
\eeq
The first constant term cancels an identical one in the initial Hamiltonian
(4), so that we may forget about these constants and concentrate only on the
resting nontrivial contributions.
Using the representation
\beq
\psi(x)-\psi(1)=\int_{0}^{\infty}dt(\exp(-t)-\exp(-xt))/(1-\exp(-t))
\eeq
and the othornormalization property of the set $\xi_{\alpha}$ we may cast
 $T^{(1)}$ in the form
\beq
T^{(1)}_{\alpha_{1},\alpha_{3}}=
2\int_{0}^{\infty}(dt/(\exp t-1))(\delta_{\alpha_{1}\alpha_{3}}-
\exp(t(1-|l_{1}|/2))g_{\alpha_{1}\alpha_{3}}(t)),
\eeq
where
\beq
g_{\alpha_{1}\alpha_{3}}(t)=4\pi\int d\nu
\xi^{\ast}_{\alpha_{1}}(\nu)\xi_{\alpha_{3}}(\nu)\cos \nu t.
\eeq
Note that (44) is not valid for ${\mbox Re}\,x=0$. Therefore  this formula
cannot be applied when the gluon orbital momentum is zero. In this case one
may use
\[ \psi(i\nu)+\psi(-i\nu)=\psi(1+i\nu)+\psi(1-i\nu),\]
which formally corresponds to changing the angular momentum to be equal to 2.

As to the second part of the kinetic energy $T^{(2)}$, we shall find out
presently that it will be cancelled by a similar contribution coming from the
monopole part of the Coulomb interaction for the angular momentum transfer
equal to zero.

\section{Monopole part of the Coulomb interaction}

\ \ \ \ Most of the contributions to the energy presented in the previous
section can hardly be further simplified and were used in the numerical
calculations as they stand. The exception is the monopole part of the Coulomb
interaction corresponding to (25) with $l=0$ (Eq. (27)). This part contains
contributions which cancel the term $T^{(2)}$ in the kinetic energy and
partially the contact interaction contribution for $l=0$. The cancellation
between the monopole Coulomb interaction and the kinetic term $T^{(2)}$ is
responsible for the scale invariance of the energy.

Explicitly the monopole term contribution is given by
\beq
 U_{\alpha_{1},\alpha_{2};\alpha_{3},\alpha_{4}}=16\pi^{2}
\int_{-\infty}^{\infty} dz_{1}\eta_{\alpha_{1}}(z_{1})\psi_{\alpha_{3}}(z_{1})
z_{1}\int_{-\infty}^{z_{1}}\psi_{\alpha_{2}}(z_{2})\eta_{\alpha_{4}}(z_{2})
+(\alpha_{1}\leftrightarrow\alpha_{4},\alpha_{2}\leftrightarrow\alpha_{3}).
\eeq
Here and in the following it is assumed that $l=0$, that is, $l_{1}=l_{3}$
and $l_{2}=l_{4}$. Introduce a function
\beq
\chi_{\alpha_{2},\alpha_{4}}(z)=\int_{-\infty}^{z}dz'
\psi_{\alpha_{2}}(z')\eta_{\alpha_{4}}(z').
\eeq
Once integrating by parts we find
\beq
\chi_{\alpha_{2},\alpha_{4}}(z)=\psi_{\alpha_{2}}(z)\xi_{\alpha_{4}}(z)-
\xi_{\alpha_{2},\alpha_{4}}(z),
\eeq
where the function $\xi_{\alpha_{2},\alpha_{4}}(z)$ with two indices, symmetric
in these, is defined as
\beq
\xi_{\alpha_{2},\alpha_{4}}(z)=\int_{-\infty}^{z}dz'
\xi_{\alpha_{2}}(z')\xi_{\alpha_{4}}(z').
\eeq
As $z\rightarrow\infty$, according to (23),
$\xi_{\alpha_{2},\alpha_{4}}(z)\rightarrow(1/4\pi)
\delta_{\alpha_{2},\alpha_{4}}$, so that
\[\chi_{\alpha_{2},\alpha_{4}}(\infty)=-(1/4\pi)
\delta_{\alpha_{2},\alpha_{4}}.\]
Having this in mind we can rewrite (47) in the form
\beq
 U_{\alpha_{1},\alpha_{2};\alpha_{3},\alpha_{4}}=16\pi^{2}
\int_{-\infty}^{\infty}
dz(\chi_{\alpha_{3},\alpha_{1}}(z)+(1/4\pi)
\delta_{\alpha_{1},\alpha_{3}})'z\chi_{\alpha_{2},\alpha_{4}}(z)
+(\alpha_{1}\leftrightarrow\alpha_{4},\alpha_{2}\leftrightarrow\alpha_{3}).
\eeq
Integrating  by parts, the integral transforms into
\beq
 -16\pi^{2}
\int_{-\infty}^{\infty} dz(\chi_{\alpha_{3},\alpha_{1}}(z)+(1/4\pi)
\delta_{\alpha_{1},\alpha_{3}})(z\chi'_{\alpha_{2},\alpha_{4}}(z)+
\chi_{\alpha_{2},\alpha_{4}}).
\eeq
The term coming from the product $\chi_{\alpha_{3},\alpha_{1}}
z\chi'_{\alpha_{2},\alpha_{4}}$ cancels the contribution
$(\alpha_{1}\leftrightarrow\alpha_{4},\alpha_{2}\leftrightarrow\alpha_{3})$
in (51) so that the monopole contribution becomes
\beq
 U_{\alpha_{1},\alpha_{2};\alpha_{3},\alpha_{4}}=-16\pi^{2}
\int_{-\infty}^{\infty} dz(\chi_{\alpha_{3},\alpha_{1}}(z)
\chi_{\alpha_{2},\alpha_{4}}(z)+(1/4\pi)
\delta_{\alpha_{1},\alpha_{3}}(z\chi'_{\alpha_{2},\alpha_{4}}(z)+
\chi_{\alpha_{2},\alpha_{4}}(z))).
\eeq

Now the idea is to substitute the functions $\chi$ in (53) by the symmetric
functions $\xi$ using relation (49). Take the the first term in the integrand
of (53). With (49) we obtain for it
\[\chi_{\alpha_{3},\alpha_{1}}\chi_{\alpha_{2},\alpha_{4}}=
\psi_{\alpha_{3}}\xi_{\alpha_{1}}\psi_{\alpha_{2}}\xi_{\alpha_{4}}
-\psi_{\alpha_{3}}\xi_{\alpha_{1}}\xi_{\alpha_{2},\alpha_{4}}
-\xi_{\alpha_{3},\alpha_{1}}\psi_{\alpha_{2}}\xi_{\alpha_{4}}
+\xi_{\alpha_{3},\alpha_{1}}\xi_{\alpha_{2},\alpha_{4}}.\]
Having in mind the subsequent symmetrization with respect to the interchange
of gluons 1 and 2, we can change $\alpha_{1}\leftrightarrow
\alpha_{2}$ and $\alpha_{3}\leftrightarrow
\alpha_{4}$ in the second term. Summed with the third term it then gives
\beq
 -\xi_{\alpha_{3},\alpha_{1}}(\psi_{\alpha_{2}}\xi_{\alpha_{4}}+
\xi_{\alpha_{2}}\psi_{\alpha_{4}}).\eeq
Recall now that $\xi_{\alpha_{2}}=(\partial+(1/2)|l_{2}|)\psi_{\alpha_{2}}$
and similarly for $\xi_{\alpha_{4}}$. Integration by parts allows  to
substitute (54) by
\beq
(\xi_{\alpha_{3}}\xi_{\alpha_{1}}-|l_{2}|\xi_{\alpha_{3},\alpha_{1}})
\psi_{\alpha_{2}}\psi_{\alpha_{4}}.
\eeq
So finally the first term in (53) leads to the following three contributions to
the monopole Coulomb energy:
\beq
\tilde{U}^{(1)}_{\alpha_{1},\alpha_{2};\alpha_{3},\alpha_{4}}=-16\pi^{2}
\int_{-\infty}^{\infty} dz\xi_{\alpha_{3},\alpha_{1}}
\xi_{\alpha_{2},\alpha_{4}},
\eeq
\beq
U^{(2)}_{\alpha_{1},\alpha_{2};\alpha_{3},\alpha_{4}}=16\pi^{2}|l_{2}|
\int_{-\infty}^{\infty} dz\xi_{\alpha_{3},\alpha_{1}}
\psi_{\alpha_{2}}\psi_{\alpha_{4}}
\eeq
and
\beq
U^{(3)}_{\alpha_{1},\alpha_{2};\alpha_{3},\alpha_{4}}=-16\pi^{2}
\int_{-\infty}^{\infty} dz(\psi_{\alpha_{3}}\xi_{\alpha_{1}}
\psi_{\alpha_{2}}\xi_{\alpha_{4}}+
\xi_{\alpha_{3}}\xi_{\alpha_{1}}
\psi_{\alpha_{2}}\psi_{\alpha_{4}}).
\eeq
Of these terms the first is  divergent in its present form. It will receive its
meaning after adding a new contributions coming from the rest of the terms
in (53). For that reason we have denoted it with a tilda.

Now for the rest of the terms in (53).  Changing the function $\chi$ by $\xi$
according to (49) we have under the integral
\[\xi_{\alpha_{2},\alpha_{4}}+z\xi'_{\alpha_{2},\alpha_{4}}=
\psi_{\alpha_{2}}\xi_{\alpha_{4}}-\xi_{\alpha_{2},\alpha_{4}}+
z\psi_{\alpha_{2}}(\partial-(1/2)|l_{2}|)\xi_{\alpha_{4}}.\]
Integration by parts transforms it into
\beq
-\xi_{\alpha_{2},\alpha_{4}}-z\xi_{\alpha_{2}}\xi_{\alpha_{4}}.\eeq
The first term can be combined with (56) to give the final part $U^{(1)}$:
\beq
U^{(1)}_{\alpha_{1},\alpha_{2};\alpha_{3},\alpha_{4}}=16\pi^{2}
\int_{-\infty}^{\infty} dz
\xi_{\alpha_{2},\alpha_{4}}((1/4\pi)\delta_{\alpha_{1},\alpha_{3}}-
\xi_{\alpha_{3},\alpha_{1}}).
\eeq
Now the integral converges due to the property (23). Putting here the explicit
form of the functions $\xi_{\alpha_{i},\alpha_{k}}$ and integrating over $z$
we obtain the term $U^{(1)}$ in its definitive form:
\beq U^{(1)}_{\alpha_{1},\alpha_{2};\alpha_{3},\alpha_{4}}=16\pi^{2}
\int dz_{1}dz_{2}(z_{1}-z_{2})\theta (z_{1}-z_{2})
\xi_{\alpha_{1}}(z_{1})\xi_{\alpha_{3}}(z_{1})
\xi_{\alpha_{2}}(z_{2})\xi_{\alpha_{4}}(z_{4}).
\eeq

 The second term in (59)
gives the last  contribution to the monopole energy
\beq U^{(4)}_{\alpha_{1},\alpha_{2};\alpha_{3},\alpha_{4}}=4\pi
\delta_{\alpha_{1},\alpha_{3}}
\int_{-\infty}^{\infty} dzz
\xi_{\alpha_{2}}\xi_{\alpha_{4}}.
\eeq
This term cancels with the contribution $T^{(2)}$ to the kinetic energy.
Indeed after the Fourier transformation to the $\nu$ space according to (33),
the factor $z$ goes into $i\partial/\partial\nu$. One can then see that (62)
gives exactly the contribution $T^{(2)}$, Eq. (41), with an opposite sign
and with gluons 1 and 2 interchanged, which is of no importance because of the
subsequent symmetrization.

The term $U^{(3)}$ cancels with the part of the
contact interaction $Q$, Eq. (30), which does not contain factors $\Delta$:
\beq
Q^{(2)}_{\alpha_{1},\alpha_{2};\alpha_{3},\alpha_{4}}=8\pi^{2}
\int_{-\infty}^{\infty} dz(
\psi_{\alpha_{3}}\xi_{\alpha_{1}}\psi_{\alpha_{2}}\xi_{\alpha_{4}}+
\xi_{\alpha_{3}}\xi_{\alpha_{1}}\psi_{\alpha_{2}}\psi_{\alpha_{4}}+
\psi_{\alpha_{3}}\psi_{\alpha_{1}}\xi_{\alpha_{2}}\xi_{\alpha_{4}}+
\xi_{\alpha_{3}}\psi_{\alpha_{1}}\xi_{\alpha_{2}}\psi_{\alpha_{4}}).
\eeq
Summed with $U^{(3)}$ this part gives
\beq (Q^{(2)}+U^{(3)}) _{\alpha_{1},\alpha_{2};\alpha_{3},\alpha_{4}}=8\pi^{2}
\int_{-\infty}^{\infty} dz
(\psi_{\alpha_{1}}\xi_{\alpha_{2}}-\xi_{\alpha_{1}}\psi_{\alpha_{2}})
(\xi_{\alpha_{3}}\psi_{\alpha_{4}}+\psi_{\alpha_{3}}\xi_{\alpha_{4}}).
\eeq
This expression is antisymmetric under the interchange of the gluons 1 and 2
and does not give any contribution to the energy.

So finally the only contributions which remain in the interaction for zero
angular momentum transfer are $U^{(1)}$, $U^{(2)}$ and the part $Q^{(1)}$ of
the contact interaction (30) which contains factors $\Delta$.

\section{Oscillator basic functions}

\ \ \ \ A natural orthonormal discrete basis for $z=\ln r^{2}$ varying from
$-\infty$ to $+\infty$ is formed by the harmonic oscillator proper functions.
Thus we choose functions $\xi_{k,l}(z)$ independent of $l$ and given by
\beq
\xi_{k}(z)=c_{k}H_{k}(z)\exp (-z^{2}/2),
\eeq
where $H_{k}$ are the Hermite polinomals and $c_{k}$ are determined by the
normalization condition (23) to be
\beq
c^{2}_{k}=1/(4\pi^{3/2}2^{k}k!).
\eeq
The  Fourier transformation to the $\nu$ space gives
\beq
\xi_{k}(\nu)=(-i)^{k}c_{k}H_{k}(\nu)\exp (-\nu^{2}/2).
\eeq
In the coordinate space the function $\eta_{k,l}(z)$ is obtained from $\xi$
by differentiation:
\beq
\eta_{k,l}(z)=(\partial-(1/2)|l|)\xi_{k}(z).
\eeq
Using the properties of $H_{k}$ we get
\beq
\eta_{k,l}(z)=2k(c_{k}/c_{k-1})\xi_{k-1}(z)-(z+(1/2)|l|)\xi_{k}(z).
\eeq
The function $\psi_{k,l}$ is obtained from $\xi_{k}$ as a solution of the
 differential equation
\beq
\xi_{k}(z)=(\partial+(1/2)|l|)\psi_{k,l}(z),
\eeq
with a boundary condition $\psi_{k,l}(-\infty)=0$. It is given by an integral
\beq
\psi_{k,l}(z)=\int_{-\infty}^{z}dz'\xi_{k}(z')\exp(-|l|(z-z')/2).
\eeq
For $k=0,1$ we find from (71):
\beq
\psi_{0,l}=\sqrt{\pi/2}\ c_{0}\exp((w^{2}-z^{2})/2)(1-\Phi(w/\sqrt{2})),
\eeq
\beq
\psi_{1,l}(z)=2(z+w)(c_{1}/c_{0})\psi_{0,l}(z)-2c_{1}\exp(-z^{2}/2),
\eeq
where $w=|l|/2-z$ and $\Phi\equiv{\rm erf}$ is the error function integral.
For $k>1$ the functions $\psi$ can be found by a recurrency relation that
follows from (71) upon integrating by parts:
\beq
\psi_{k+1,l}=|l|(c_{k+1}/c_{k})\psi_{k,l}+2k(c_{k+1}/c_{k-1})\psi_{k-1,l}-
2(c_{k+1}/c_{k})\xi_{k}.
\eeq

With this set of functions the potential part of the energy was calculated
numerically. As to the kinetic part, the function $g$ entering (45) can be
found analytically. For transition $k,l\rightarrow k',l$ it is equal to zero if
$k+k'$ is odd. For even $k+k'=2s$ and $k\geq k'$
\beq
g_{kk'}(t)=4\pi^{3/2}(-1)^{d}c_{k}c_{k'}\exp(-t^{2}/4)\sum_{p=0}^{k'}
2^{p}p!C_{k}^{p}C_{k'}^{p}(-t^{2})^{s-p},
\eeq
where $2d=k-k'$.

We finally note that the presented set of basic functions can be trivially
generalized to include a scaling factor $a$ in the $z$ space by choosing as
basic functions
\beq
\xi_{k}^{(a)}(z)=\sqrt{a}\xi_{k}(az).
\eeq
Other functions can then easily be found as
\beq
\xi_{k}^{(a)}(\nu)=(1/\sqrt{a})\xi_{k}(\nu/a),
\eeq
\beq
\eta_{k,l}^{(a)}(z)=a\sqrt{a}\eta_{k,l/a}(az)
\eeq
and finally
\beq
\psi_{k}^{(a)}(z)=(1\sqrt{a})\psi_{k,l/a}(az).
\eeq
Putting these expressions into the obtained formulas for the energy and
rescaling the variables $z$ or $\nu$ we find that the  new value of
the potential energy is obtained by dividing all angular momenta by $a$ and
dividing the resulting energy by $a$. As to the kinetic energy the change
reduces to substituting $at$ instead of $t$ in the function $g$, Eq. (45).

\section {Results for $n=2,3$ and conclusions}

\ \ \ \ We applied this formalism to the cases $n=2$ (the BFKL pomeron) and
$n=3$ (the odderon) to study its convergence.

The energy matrix ${\cal E}_{\alpha_{1},\alpha_{2},\alpha'_{1},\alpha'_{2}}$
has been calculated numerically for a chosen set of basic functions described
in the previous section. After proper symmetrization in two- or three-gluon
states its lowest eigenvalue has been determined, which gives an upper limit
on the exact pomeron or odderon energy according to Eq. (11). To study the
minimal energy only states with the total angular momentum equal to zero have
been included.

The selected set of basic one-gluon functions is characterized by the maximal
value of the angular momentum included $l_{max}$ and numbers of radial
functions included for each wave. As calculations show, best results are
obtained when one raises $l_{max}$ and the number of radials in all waves
simultaneously. So we present here the results for the case when the number
of radials $r$ is the same for all angular momenta and is equal to the number
of angular momenta included $r=l_{max}+1$. Such a set of functions is thus
characterized by a single parameter $r$. With a growth of $r$ the number of
states $N$ rises very rapidly. For two gluons $N_{2}=r(r^{2}-r/2-1/2)$ and so
rises as $r^{3}$. For three gluons the rise is still  steeper.

In the present calculations the number $r$ was limited by 6 for two gluons
and by 5 for three gluons. Correspondingly the total number of basic states
included was taken up to 201 for two gluons and 1335 for three gluons.

The results of the calculations of the ground state energies of the BFKL
pomeron $\epsilon_{2}$ and the odderon $\epsilon_{3}$ are presented in the
Table  for different values of $r$.
One observes that the obtained energies are still rather far from the exact
value for the pomeron and the upper limit for the odderon obtained in [5].
Thus the convergence of the method is rather slow. The Table also reveals
 that the odderon energy is essentially
larger than the pomeron one for a given $r$. So our results confirm
that, in all probability, the odderon intercept is lower than that of the BFKL
pomeron.

To be more quantitative one can estimate the precision of the variational
results by comparing the calculated correlation energy with its exact value
known for the BFKL pomeron. Subtracting $E_{0}$ from the Hartree-Fock energy
we find $\epsilon_{2}^{cor}=3.732$. For the maximal value $r=6$ the
calculated correlation energy is 1.991, which constitutes $\sim 53\%$ of the
exact value. With $r=5$ the correlation energy is $1.871$, that is $\sim50\%$
of the exact value. If one boldly assumes that for the odderon  the correlation
energy calculated with $r=5$ also constitutes $50\%$ of the exact value, then
one finds this exact value to be 1.256 and the absolute energy
$\epsilon_{3}=-0.3$, that is, $E_{odd}=-0.45$, which does not contradict
 [5].

 More sophisticated estimates can be attempted by studying the dependence
of the calculated energy on $r$ and
 extrapolating for higher values of $r$. We have chosen a 4-parameter fit
\beq
\epsilon(r)=\epsilon(\infty)+a\exp({-\alpha}\ln^{\beta}r).
\eeq
For the BFKL pomeron $\epsilon (\infty)$ is known. It turns out that the
values of $\epsilon_{2}(r)$ given in the Table are well described by (80) with
the choice of parameters
\[a=6.066,\ \alpha=0.925,\ \beta=0.515.\]
According to this fit further improvement of the value for $\epsilon_{2}$
requires very high values of $r$. E.g., to achieve $\epsilon_{2}(r)=-2.0$ one
has to raise $r$ to $\sim 100$.  The corresponding numbers of
two-gluon states are enormous and hardly possible to include.

The analogous fit for the three-gluon case determined from energy values for
$r=2,3,4,5$ has the parameters
\[
\epsilon(\infty)=-0.389,\ a=2.472,\ \alpha=1.04,\ \beta=0.36.
\]
However this set of parameters is rather unstable: a small change in the
energy values used causes rather large changes in the value of
$\epsilon(\infty)$. If we take for $\beta=1/2$, as evidently favoured by the
two gluon case, then we obtain
\[\epsilon(\infty)=-0.195,\ a=1.772  \ \alpha=0.957,\ \beta=0.5,\]
with the value $\epsilon_{3}(2)=0.603$ slightly smaller than the calculated one
$0.605$ although within the calculational errors of the order
of $\pm 0.003$.  In view of this we can only give a crude estimation for
the upper limit for the odderon energy  from
our calculations,
\[\epsilon_{3}<-0.2\div-0.4,\]
which according to (11) translates into the estimate for $E_{odd}$
\[E_{odd}<-0.3\div-0.6,\]
in agreement with [5].

To conclude, the calculations for two and three gluons show that the developed
method can be applied to investigate the intercept of symmetric
configurations although the convergence is slow, evidently, due to a very
singular character of the gluonic wave function. It seems realistic to obtain
around 50\% of the correlation energy with this approach, which may serve to
estimate  intercepts for multigluon configurations relative to the BFKL one. We
are trying to achieve  better results by including more basic functions and
also selecting configurations which give the dominant contribution.
Calculations for $n=4$ gluons are also in progress.

\section{Acknoledgements}

\ \ \ \ The authors are grateful to the General Direction
of the Scientific and Techical Investigation (DGICYT) of Spain and to the
Xunta de Galicia for financial support.

\newpage
{\Large\bf References}\\

\noi[1] E. A. Kuraev, L. N. Lipatov and V. S. Fadin, Sov. Phys. JETP {\bf 44}
(1976) 433; {\bf 45} (1977) 199;
Ya. Ya. Balitzky and L. N. Lipatov, Sov. J. Nucl. Phys. {\bf 28} (1978) 822.

\vspace{0.2cm}
\noi[2] E. M. Levin, preprint FERMILAB-CONF-94/068-T (1994).

\vspace{0.2cm}
\noi[3] L. D. Faddeev and G. P. Korchemsky, preprint ITP-SP-14
(hep-th/9404173).

\vspace{0.2cm}
\noi[4] P. Gauron, L. N. Lipatov and B. Nicolescu, Phys. Lett. {\bf B260}
(1991) 407.

\vspace{0.2cm}
\noi[5] P. Gauron, L. N. Lipatov and B. Nicolescu, Z.Phys. {\bf C63} (1994)
253.

\vspace{0.2cm}
\noi[6] L. N. Lipatov, Phys. Lett. {\bf B251} (1990) 284.

\vspace{0.2cm}
\noi[7] M. A. Braun, Phys. Lett. {\bf B337} (1994) 354.

\vspace{0.2cm}
\noi[8] L. N. Lipatov, Sov. Phys. JETP {\bf 63} (1986) 904;
J. Bartels, Nucl. Phys. {\bf B175} (1980) 365; J. Kwiecinsky
and M. Praszalowicz, Phys. Lett. {\bf B94} (1980) 413.

\vspace{0.2cm}
\noi[9] M. A. Braun and C. Pajares, Phys. Lett. {\bf B287} (1992) 154; Nucl.
Phys. {\bf B390} (1993) 542; 559.

\newpage
{\Large\bf Table }
\vspace{0.5cm}\begin{quotation}
Calculated values of the ground state energy per gluon (multiplied by 2,
Eq. (11)) for the pomeron ($\epsilon_{2}$) and odderon ($\epsilon_{3}$)  with
different numbers
$r$ of radial functions and angular momenta included.
\end{quotation}
\vspace{0.2cm}
\begin{center}
\begin {tabular}{|c|c|c|}  \hline
  $r$ & $\epsilon_{2}$ & $\epsilon_{3}$\\\hline
   1  &  0.968        &  0.968     \\\hline
   2  &  0.022        &  0.605     \\\hline
   3  &$-0.475$       &  0.454     \\\hline
   4  &$-0.743$       &  0.379     \\\hline
   5  &$-0.912$       &  0.331     \\\hline
   6  &$-1.032$       &             \\\hline
      &                &             \\\hline
$\infty$& $-2.773$     &             \\\hline
\end{tabular}
\end{center}
\end{document}